\documentclass[a4paper]{jpconf}
\usepackage{graphicx}
\usepackage[compress,sort]{cite}
\begin{document}
\title{Toward parton equilibration with improved parton interaction matrix elements}

\author{Bin Zhang}

\address{Department of Chemistry and Physics,
Arkansas State University,
P.O. Box 419,
State University,
AR 72467-0419, U.S.A.
}

\ead{bzhang@astate.edu}

\begin{abstract}
The Quark-Gluon Plasma can be produced in high energy heavy ion 
collisions and how it equilibrates is important for the extraction 
of the properties of strongly interacting matter. A radiative 
transport model can be used to reveal interesting characteristics 
of Quark-Gluon Plasma thermalization. For example, screened 
parton interactions always lead to partial pressure isotropization. 
Systems with different initial pressure anisotropies evolve toward 
the same asymptotic evolution. In particular, radiative processes 
are crucial for the chemical equilibration of the system. Matrix 
elements under the soft and collinear approximation for these 
processes, as first derived by Gunion and Bertsch, are widely used. 
A different approach is to start with the exact matrix elements for
the two to three and its inverse processes. General features of 
this approach will be reviewed and the results will be compared 
with the Gunion-Bertsch results. We will comment on the possible 
implications of the exact matrix element approach on Quark-Gluon 
Plasma thermalization. 
\end{abstract}

\section{Introduction}
Many interesting discoveries have been made in the quest for
the understanding of nuclear matter under extreme 
conditions \cite{Adcox:2004mh,Arsene:2004fa,Back:2004je,
Adams:2005dq,Kovtun:2004de,McLerran:2007qj,Abelev:2009ac,
Alver:2010gr,Abelev:2010rv}.
Relativistic heavy ion collisions are particularly useful
in creating the phase of matter called the Quark-Gluon
Plasma (QGP) \cite{Heinz:2008tv}. During these collisions, 
radiative processes 
are important for the thermalization of the Quark-Gluon Plasma. 
Xu and Greiner first introduced the stochastic method into 
relativistic transport model simulations \cite{Xu:2004mz}. 
This enabled the microscopic study of the Quark-Gluon
Plasma with particle number changing processes. We also 
developed a similar algorithm for the dynamical study of 
the Quark-Gluon Plasma. In the following, we will illustrate
some interesting features of the thermalization of a
gluon system. For the simulation, the perturbative 
Quantum-Chromo-Dynamics (pQCD) cross section regulated by 
a Debye screening mass will be used for the two to two
process. It is proportional to the strong interaction 
coupling constant (fine structure constant), $\alpha_s$, squared.
The screening mass squared is proportional to $\alpha_s$, 
inversely proportional to the cell volume, and proportional to 
the sum of inverses of particle momenta. This helps avoid 
many conceptual and technical problems associated with 
large cross sections in dense media. The  two to three cross 
section is taken to be $50\%$ of the two to two cross section. 
This is in line with a more sophisticated calculation 
by Xu and Greiner \cite{Xu:2004mz}. The outgoing particles 
will be taken to be isotropic. The three to two reaction 
integral is determined by detailed balance and in this case is 
directly proportional to the two to three cross section. 

With this setup, various aspects of thermalization of 
gluons in a box can be studied. For example, for a system having
2000 gluons initially with a temperature of 1 GeV inside a box 
of dimensions of $5\times 5\times 5$ fm$^3$,
the three to two rate per unit volume approaches that
for the two to three process from below, and the particle energy 
distribution relaxes to that in thermodynamical equilibrium 
(kinetic and chemical equilibrium). More details 
can be found in \cite{Zhang:2009dk} and \cite{Zhang:2011tx}. 
In the following, we will apply the above radiative transport
model to the study of the pressure anisotropy and
energy density evolutions in relativistic heavy ion collisions.
We will then look at the two to three and its inverse processes 
with exact matrix elements for a more realistic description of
relativistic parton dynamics. Finally, a summary will be given 
together with speculations on the implications of the inelastic 
processes with improved matrix elements.

\section{Pressure anisotropy and energy density evolutions}
\begin{figure}[h]
\centering
\includegraphics[width=18pc]{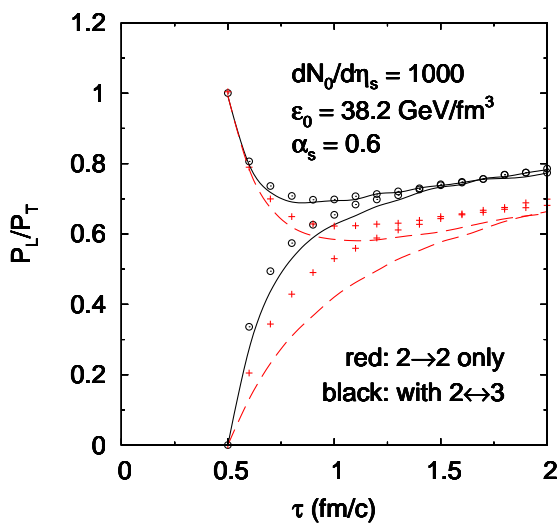}
\hspace{2pc}%
\raisebox{4pc}{
\begin{minipage}[b]{16pc}
\caption{\label{fig_plopt1}Pressure anisotropy evolutions in the central
cell in central heavy ion collisions. The lines are for evolutions
from thermal initial conditions while the points are for idealized
Color-Glass-Condensate initial conditions. The dashed lines and 
the pluses have elastic collisions only while the solid lines 
and circles include also lowest order inelastic collisions.}
\end{minipage}
}
\end{figure}

The initial stage of a relativistic heavy ion collision
is dominated by gluons. We will focus on the thermalization
of a gluon system in the central cell in central collisions.
In this case, kinetic equilibration can be characterized by 
the pressure anisotropy (i.e., the longitudinal to 
transverse pressure ratio, $P_L/P_T$) \cite{Zhang:2008zzk,
Huovinen:2008te,El:2009vj,Denicol:2010xn,Zhang:2010fx}. 
If the system is in 
equilibrium, the pressure anisotropy equals 1; if the pressure 
anisotropy is different from 1, the system is not in equilibrium.
Fig.~\ref{fig_plopt1} shows the pressure anisotropy
evolutions for initial conditions similar to those
in central Au+Au collisions at the Relativistic Heavy
Ion Collider (RHIC). We can start by looking at the 
isotropic initial condition with only elastic collisions. 
Even though the initial condition is isotropic, the longitudinal
expansion makes it decrease with (proper) time. Expansion
dominates the initial period until collisions take over
and the pressure anisotropy begins to increase
toward isotropy. If the two to three and three to two
processes are also included, more thermalization can be achieved.
If the initial condition is transverse instead of isotropic,
the pressure anisotropy increases as a result of particle
collisions. It approaches that starting from an isotropic
initial condition. If the initial condition is
Color-Glass-Condensate like instead of thermal, more
thermalization can be obtained when only elastic collisions
are included. If inelastic processes are also allowed, the 
evolutions are almost identical to those from thermal initial 
conditions due to fast exponentiation of the momentum spectra. 
More discussions can be found in \cite{Zhang:2010fx}.

The bulk properties of the central cell can be described by the energy
density, the longitudinal and transverse pressures. For a gluon system,
they are related by the zero trace of the energy momentum tensor. 
Therefore, there are only two independent variables. The energy 
density thus provides
additional information relative to the pressure anisotropy.
In particular, the early stage energy evolution reflects the 
initial anisotropy and the late time evolution is determined 
by parton interactions. Further discussions on its implications 
can be found in Ref.~\cite{Zhang:2009dk}.

\section{Improved matrix elements for inelastic gluonic processes}

To get a better description of thermalization, more realistic 
radiative matrix elements need to be implemented. 
The exact two to three matrix element was first studied in
the late 70's \cite{Gottschalk:1979wq,Berends:1981rb}.
The matrix element modulus squared (averaged over initial
internal degrees of freedom and summed over final)
can be expressed in a very symmetric form as:
\begin{equation}
|M_{gg\rightarrow ggg}|^2=\frac{g^6N_c^3}{2(N_c^2-1)}
\frac{\sum (ij)^4\sum (ijklm)}{\prod (ij)}.
\end{equation}
In the above equation, the strong interaction coupling constant
$g$ is related to $\alpha_s$ by $\alpha_s=g^2/(4\pi)$. The
number of colors is $N_c=3$. The sums and the product are over 
all distinct permutations of the set of particle labels 
$\{1,2,3,4,5\}$.
$(ij)=p_i\cdot p_j$ is the product of the four-momenta of particles
$i$ and $j$, and the string $(ijklm)=(ij)(jk)(kl)(lm)(mi)$. This
symmetric form puts all particles on an equal footing. The
denominator comes from particle propagators, and we will regulate
these propagators by the Debye screening mass squared, $\mu^2$.

It is instructive to look at some representative numbers. We will 
set $\alpha_s=0.47$. For about 300 MeV 
temperature, we have approximately $\mu^2=10$ fm$^{-2}$, 
and $s=4$ GeV$^2$ for the center-of-mass energy squared.
Then the calculated two to two cross section is 
$\sigma_{22}=0.312$ fm$^2$, and the two to three 
cross section is $\sigma_{23}=0.0523$ fm$^2$. This
gives a ratio of $\sigma_{23}/\sigma_{22}=0.168$, much smaller
than $50\%$. One can also look at the small coupling limit by
taking $\alpha_s=0.3$. This leads to a change in $\mu^2$ to
6.38 fm$^{-2}$. The calculated $\sigma_{22}=0.199$ fm$^2$, and
$\sigma_{23}=0.0504$ fm$^2$. Their ratio is also much smaller
than $50\%$.

\begin{figure}[h]
\centering
\includegraphics[width=20pc]{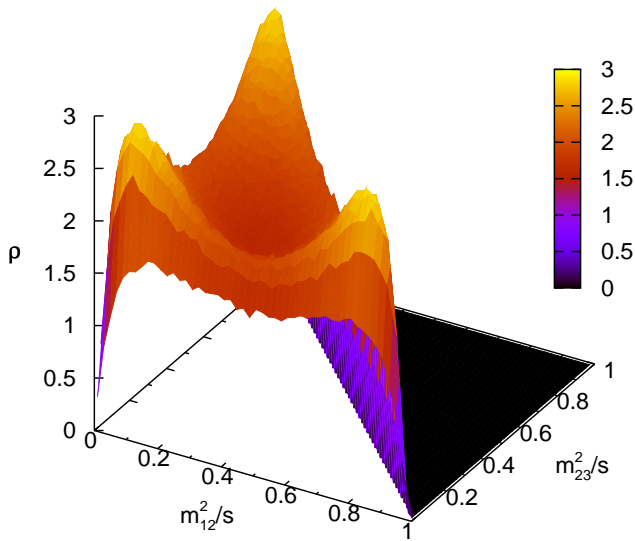}
\hspace{1pc}%
\raisebox{6pc}{
\begin{minipage}[b]{15pc}
\caption{\label{fig_dalitz1}Normalized Dalitz plot for the outgoing
particles in the $gg\rightarrow ggg$ process.}
\end{minipage}
}
\end{figure}

In addition to the total cross section which determines the 
collision rate, it is important to see how
different the outgoing particle distribution is from 
the isotropic distribution. This can be achieved by
studying the normalized Dalitz plot. 
It gives the distribution of outgoing particles as a function
of $m_{12}^2/s$ and $m_{23}^2/s$ where $m_{ij}$ is the
invariant mass of the subsystem composed of particles $i$ and $j$. 
If the outgoing particles are isotropically distributed,
the Dalitz plot is flat at $\rho=2$. Fig.~\ref{fig_dalitz1} shows
the distribution when $\mu^2=10$ fm$^{-2}$ and $s=4$ GeV$^2$.
The three peaks come from the soft and collinear
singularities. With the Debye mass regularization, the
outgoing particle distribution is not far from isotropic.

\begin{figure}[h]
\centering
\begin{minipage}{16pc}
\includegraphics[width=16pc]{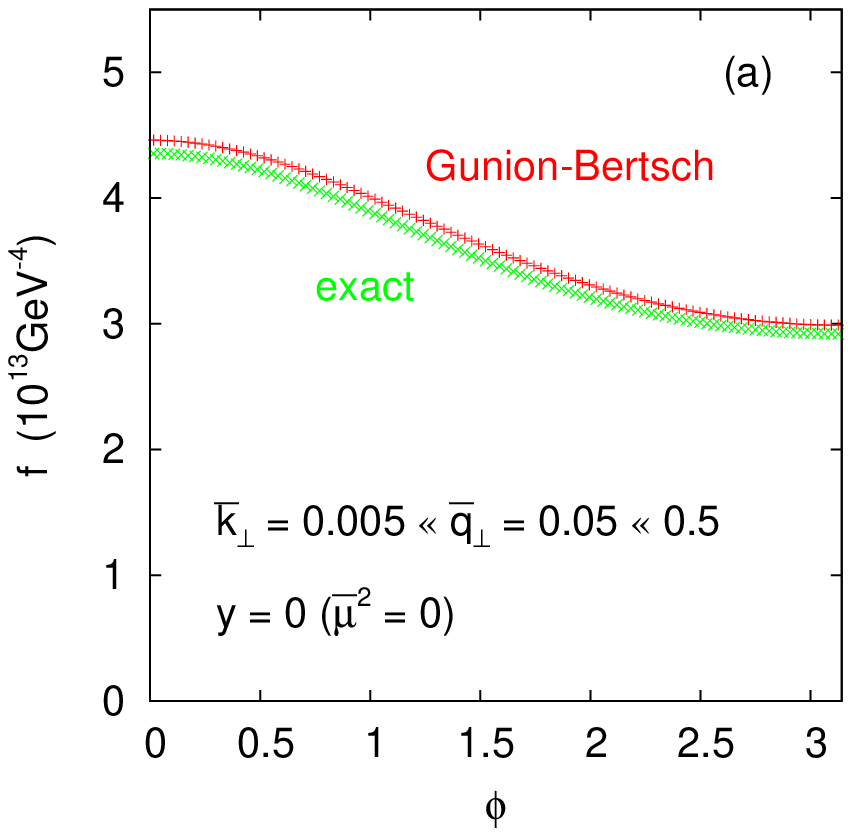}
\end{minipage}
\hspace{1pc}%minipages won't overlap
\begin{minipage}{16pc}
\includegraphics[width=16pc]{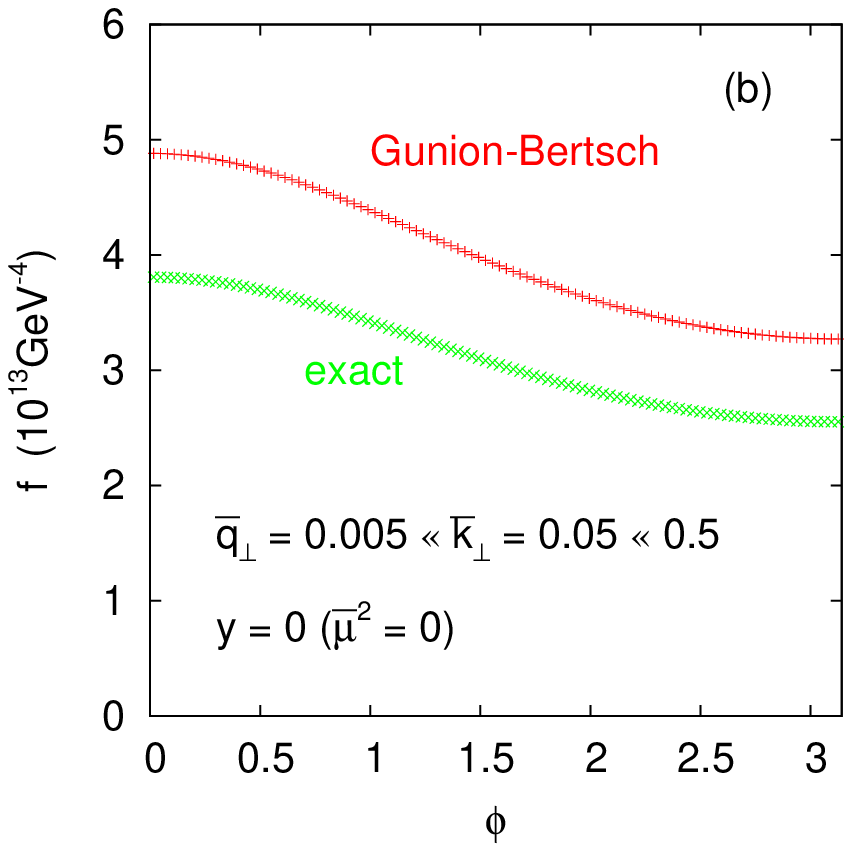}
\end{minipage}
\begin{minipage}{16pc}
\includegraphics[width=16pc]{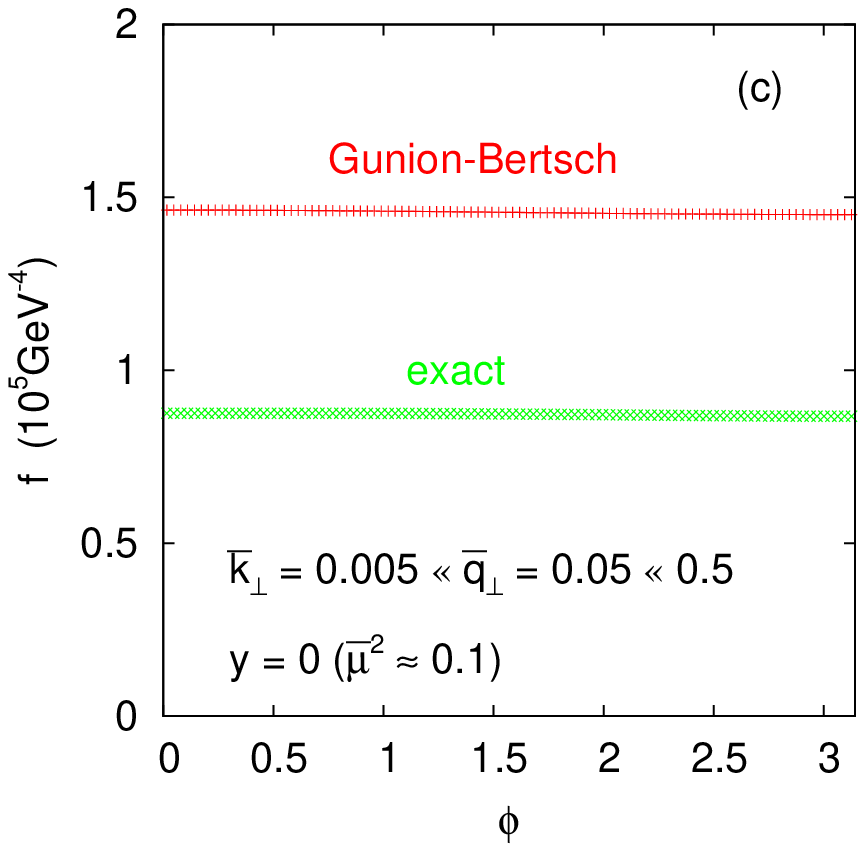}
\end{minipage}
\hspace{1pc}%
\raisebox{2pc}{
\begin{minipage}[b]{16pc}
\caption{\label{fig_exactGB1}Sums of weighted matrix elements
modulus squared as functions of the azimuthal angle $\phi$. 
}
\end{minipage}
}
\end{figure}

The above $\sigma_{23}/\sigma_{22}$ is smaller than the typical
value from the Gunion-Bertsch formula. As the Gunion-Bertsch
formula is an approximation of the exact 
formula \cite{Gunion:1981qs}, comparisons
between particle distributions from these two formulas can 
illustrate the origins of the differences. The Gunion-Bertsch 
formula is conventionally expressed as a function of the 
transverse momentum transfer $\vec{q}_\perp$ and the transverse 
momentum of the radiated gluon $\vec{k}_\perp$ as 
\begin{equation}
|M_{gg\rightarrow ggg}^{GB}|^2=\frac{9g^4s^2}{2(q_\perp^2+\mu^2)^2}
\frac{12g^2q_\perp^2}{k_\perp^2((\vec{k}_\perp-\vec{q}_\perp)^2+\mu^2)}.
\end{equation}
We will use the Debye screening mass squared to regulate
the singularity when $k_\perp\rightarrow 0$. Fig.~\ref{fig_exactGB1}
shows some comparisons for $s=4$ GeV$^2$. 
The horizontal axis is $\phi$, the azimuthal 
angle between $\vec{k}_\perp$ and $\vec{q}_\perp$. The vertical 
axis is $f(q_\perp,k_\perp,y,\phi)=
\sum_{y'_{1a},y'_{1b}}|M|^2/|\partial F/\partial y'_1|_{F=0}$.
In the above expression, $y$ is the rapidity of the radiated
gluon. The rapidity of the outgoing gluon that acquires 
the transverse momentum transfer is $y'_1$ , and
$y'_{1a}$ and $y'_{1b}$ are the roots of $F=0$ where
$F$ is the four-momentum squared of the particle other 
than the radiated and the transverse momentum transferred
ones. Only results for $y=0$ are shown here. 
The singularity is not regulated for the top panels.
In other words, $\bar{\mu}^2=0$ for these two cases. 
In the following, unless stated 
otherwise, barred symbols are for variables reduced by $\sqrt{s}$.
We see that when $\bar{k}_\perp$ is much smaller than 
$\bar{q}_\perp$ and they are both much smaller than the kinematics 
limit ($0.5$), the Gunion-Bertsch result is only slightly 
larger than the exact result. However, if $\bar{q}_\perp$ is much 
smaller than $\bar{k}_\perp$, the Gunion-Bertsch result can be 
higher than the exact by $25\%$. 
Comparison of the left panels shows that the regulator can
significantly reduce the magnitudes and the Gunion-Bertsch result 
can be higher than the exact by as much as $50\%$. 
We also looked at other kinematic regions and 
the two do not always agree. 

\begin{figure}[h]
\centering
\includegraphics[width=34pc]{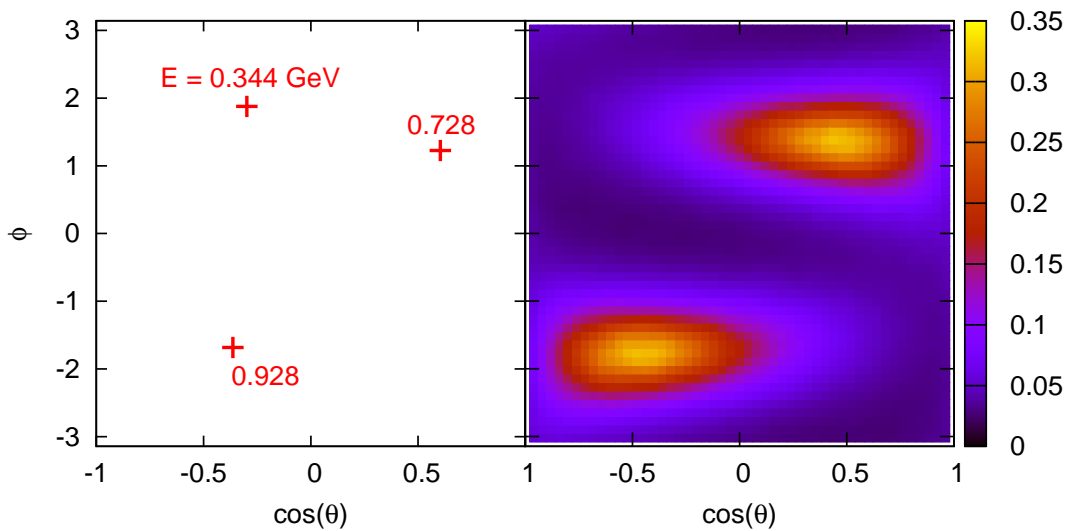}
\caption{\label{fig_3to2a}The incoming (left) and outgoing (right) 
particle distributions for a three to two process. The numbers in the
left panel are for particle energies. The outgoing particle
distribution is normalized to 1.}
\end{figure}

\begin{figure}[h]
\centering
\includegraphics[width=34pc]{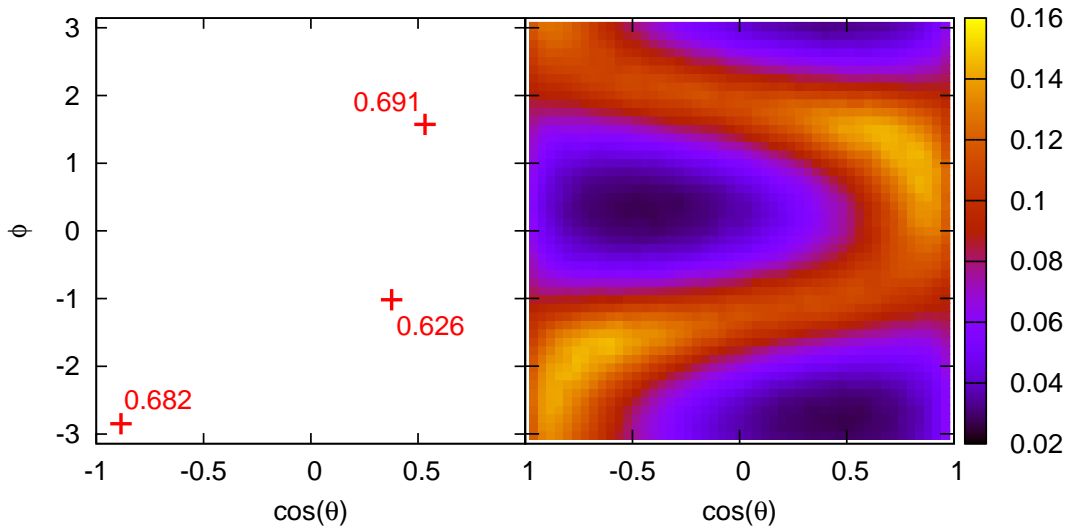}
\caption{\label{fig_3to2b}Like Fig.~\ref{fig_3to2a} but for
three different incoming particles.}
\end{figure}

Besides the two to three process for particle production, 
its inverse
process is also important in the thermalization of a gluon system.
In the following, we will calculate the reaction integral and 
look at the outgoing particle distribution.  Fig.~\ref{fig_3to2a}
gives an example of the initial and final distributions of 
particles as functions
of $\cos(\theta)$ and $\phi$ in the center-of-mass frame. Notice
that there is a soft particle in the initial state and the final
distribution is for the first outgoing particle with the other 
balancing the momentum of the first one. The
interactions are specified by $\alpha_s=0.47$ and $\mu^2=10$ 
fm$^{-2}$. The calculated reaction integral is $I_{32}=6.84$ fm$^2$,
close to the estimate (6.19 fm$^2$) if the matrix 
element is isotropic. The outgoing particle distribution 
has a two-peak structure strongly affected by the two ``hard" particles 
and the soft particle appears to be absorbed. Fig.~\ref{fig_3to2b} 
provides another example. In this case, the three incoming particles have 
about the same energy. The reaction integral is
$I_{32}=4.85$ fm$^2$, smaller than that estimated
from the isotropic matrix element formula. The outgoing particle
distribution is a ring determined by the incoming particles.
This is also very different from the uniform distribution when
the matrix element is isotropic.

\section{Summary and speculations}
Radiative transport plays an important role in momentum 
space exponentiation and particle production in relativistic
heavy ion collisions.
The above calculations of typical two to three and two to two
cross sections indicate that elastic collisions may be more
important in thermalization than expected from the 
Gunion-Bertsch formula based calculations. Xu and Greiner 
obtained the specific shear viscosity (shear viscosity to
entropy density ratio) and showed that it approaches the 
conjectured quantum limit at large $\alpha_s$ \cite{Xu:2007ns}. 
Their calculations
were based on the Gunion-Bertsch formula. As the exact
formula based radiative cross section can be much smaller,
the specific shear viscosity may be much larger than the 
quantum limit. If so, this will be in qualitative agreement 
with calculations by Chen {\it et al.} based on the exact matrix 
elements \cite{Chen:2009sm,Chen:2010xk}. 
The above outgoing particle distributions show that the two 
to three process is not far from isotropic while the three to two 
is not quite close. Hence specific viscosity calculations need 
to be explicitly carried out to see the difference.

The comparison of the exact and Gunion-Bertsch matrix elements
shows big differences in key kinematic regions. The 
Gunion-Bertsch formula was regulated with the screening mass.
This is equivalent to replacing the original theta function
used by Xu and Greiner by a Lorentzian. In other words, the 
screened propagator and the theta function are two different
ways of saying the same thing, i.e., soft interactions are
limited by the Landau-Pomeranchuk-Migdal(LPM) effect. It is 
also interesting to compare the formation time and the
mean free path based on the above calculations. 
It turns out that they are on the same order for 
processes that are important for thermalization. 
This is very different from processes involving jets
where the coherent length can be much longer 
than the mean free path \cite{Wong:2011nd}. Therefore, thermalization
can be much more sensitive to the space-time evolution of the
hot and dense nuclear medium compared to jets. The above 
picture may not be limited to gluon and light quark 
processes only. Heavy quark equilibration may also 
benefit strongly from elastic processes and some 
other quasi-elastic processes such as the meson dissociation 
process \cite{Adil:2006ra}. Hopefully calculations in 
the near future will be able to clarify some of these 
questions.

\ack
We would like to thank W. A. Wortman for participation in part of
the work and the Parallel Distributed Systems Facilities of the
National Energy Research Scientific Computing Center for providing 
computing resources. This work was supported by the U.S. National 
Science Foundation under grant No. PHY-0970104.

\end{document}